\documentclass[prd,aps,floatfix,nofootinbib,preprint,tightenlines ]{revtex4}
\usepackage{dcolumn}% Align table columns on decimal point
\usepackage{amsmath}
\usepackage{latexsym}
\usepackage{graphicx}
\usepackage{bm}
\def\Tr{{\rm Tr}\,}

\newcommand{\bee}{\begin{equation}}
\newcommand{\ee}{\end{equation}}
\newcommand{\beea}{\begin{eqnarray}}
\newcommand{\eea}{\end{eqnarray}}

\begin{document}
%%%%%%%%%%%%%%%%%%%%%%%%%%%%%%%%%%%%%%%%%%%%%%%%%%%%%%%%%%%%%%%%%%%%%%
\title{
Oblique correction in a walking lattice theory
}
\author{Thomas DeGrand}
\email{thomas.degrand@colorado.edu}
\affiliation{Department of Physics,
University of Colorado, Boulder, CO 80309, USA }

\begin{abstract}
I compute the difference of vector and axial vector current correlators in the 
weak coupling phase of (lattice-regulated) SU(3) gauge theory with two flavors
of symmetric-representation dynamical fermions. This is a walking theory
at the bare parameter values chosen for the 
simulation. Otherwise, it is not a conventional technicolor candidate.
The correlator difference shows 
scaling behavior  in the fermion mass, and vanishes in the fermion zero mass limit.
Consequences for the phenomenology of similar systems which
might be candidates for beyond Standard Model physics are discussed.
I check my methodology against ordinary QCD, by computing the Gasser-Leutwyler 
coefficient $\overline L_{10}$
and the charged - neutral pion mass difference from an approximate parameterization of
the correlator.
\end{abstract}

%\pacs{11.15.Ha,  12.60.Nz}
\maketitle

%%%%%%%%%%%%%%%%%%%%%%%%%%%%%%%%%%%%%%%%%%%%%%%%%%%%%%%%%%%%%%%%%%%%%
\section{Introduction and background}
%%%%%%%%%%%%%%%%%%%%%%%%%%%%%%%%%%%%%%%%%%%%%%%%%%%%%%%%%%%%%%%%%%%%%
We hope to discover new physics at the Large Hadron Collider. But even if a direct observation of
physics beyond the Standard Model is difficult, it still may be possible to detect new physics 
through its effect on Standard Model processes, via precision  electroweak measurements.
One classic observable is the S-parameter of Peskin and Takeuchi\cite{Peskin:1991sw}.
In the simple system I will analyze, it is related to the difference between the
momentum space vector and axial current correlators at zero momentum
\beea
\Pi_{\mu\nu}(q) &=& \int {d^4 q} \exp(iqx) \langle J_\mu^L(x) J_\nu^R(0)\rangle \nonumber \\
   &\equiv & (q^2 \delta_{\mu\nu} - q_\mu q_\nu)\Pi_T^{LR}(q^2) + q_\mu q_\nu\Pi_L^{LR}(q^2) . \nonumber .\\
\label{eq:continPi}
\eea
The S-parameter is proportional to the limiting value of $d(q^2\Pi_T^{LR}(q^2))/dq^2$ at small $q^2$,
after Goldstone boson effects are subtracted.

Recently, many groups have begun to use lattice methods
to study candidate beyond-Standard Model systems which replace the fundamental
Higgs field by new, nonperturbative physics \cite{reviews}.
These models may be realizations of technicolor (for reviews, see Ref.~\cite{Hill:2002ap}),
or they may correspond to ``hidden sector''\cite{Strassler:2006im}  or ``unparticle'' 
systems \cite{Georgi:2007ek}
where the new physics is approximately conformal.
To decide whether any particular model is viable, it is
necessary to compute observables. This paper describes the measurement of $\Pi^T_{LR}$  for one
candidate theory, $SU(3)$ gauge fields coupled to $N_f=2$ flavors of fermions in the
symmetric (sextet) representation.
Before lattice simulations began, this model was one of many candidate theories for
walking technicolor  \cite{Sannino:2004qp,Hong:2004td,Dietrich:2006cm,Ryttov:2007sr}.
 Previous lattice simulations of this system include
Refs.~\cite{Shamir:2008pb,DeGrand:2008kx,DeGrand:2009hu,DSS,Sinclair:2009ec,Fodor:2009ar}.
In this system, $\Pi_{\mu\nu}(q)$ involves the correlation function of two fermionic ($\psi$)
 bilinears.
In my conventions $J_\mu^L= \bar \psi \gamma_\mu \frac{(1-\gamma_5)}{2} \psi$ and
  $J_\mu^R= \bar \psi \gamma_\mu\frac{(1+\gamma_5)}{2} \psi$ .

The specific lattice calculation is done at a set of bare parameters at which the gauge coupling
runs slowly or ``walks'': the theory is approximately conformal in the zero fermion mass limit.
Tuning the fermion mass away from zero explicitly breaks conformal symmetry.
What happens to $\Pi_T^{LR}(q^2)$ in this situation is, as far as I know, unexplored.
Unfortunately, this system is not a candidate for conventional technicolor.
At the parameter value where I did the simulations,
 it  shows no sign of spontaneous chiral symmetry breaking.

There is some discussion in the literature of $\Pi_T^{LR}(q^2)$ for
 technicolor candidates which have slowly running or ``walking'' couplings, or which are
conformal \cite{Luty:2008vs}.
 In these systems, conformal symmetry is is usually
 broken, either
explicitly or through the coupling of the new physics sector to the Standard Model.
Here, the new
 physics at very short distance which generates a fermion mass,
is  a bare fermion mass in the lattice action.
 The question is then, what happens to $\Pi^T_{LR}$ as the mass is tuned to zero.
 The expectation is that $\Pi^T_{LR}$ falls to zero in that limit.
A second expectation often seen in the Beyond Standard Model literature is that
$\Pi^{LR}$ can be computed by saturation by a few light resonances in the appropriate channels, by
tuning their masses and couplings. I will test these expectations.

By ``slowly running'' I mean that a suitably defined running coupling constant shows
small variation, over the range of
 length scales accessible to a lattice simulation in finite volume at  a fixed value of its 
bare couplings.
In Ref.~\cite{DSS}, this is done using
a lattice version of the background field method, in which the system size $L$ represents the
 scale at which
a running coupling $g^2(L)$ is measured. (It is called the Schr\"odinger functional method.
Part of the extensive literature of the Schr\"odinger functional include
Refs.~\cite{Luscher:1993gh,Jansen:1998mx,Sint:1995ch,DellaMorte:2004bc}.)
It was observed \cite{DSS} that the coupling ran more slowly than perturbation theory predicted.
 At one loop, this is
\bee
\frac{1}{g^2(sL)} = \frac{2b_1}{16\pi^2}\log sL + \rm{constant} .
\ee
where $b_1=13/3$ for this theory. For a scale factor $s=2$, this is a change in 
$1/g^2(sL)$ of
about 0.038. compared to the measured value at this simulation's bare coupling
of $1/g^2(L)$ of about 0.38.

This is not what happens in ordinary QCD. With two flavors of fundamental representation fermions,
$b_1=29/3$ and in perturbation theory the change in the inverse coupling over a scale factor of two is
about 0.084. In practice, at coupling values used in standard QCD simulations, the change is much
 greater. (Ref.~\cite{DellaMorte:2004bc} compares their Schr\"odinger functional coupling
to perturbation theory.)
In fact, the whole framework of high precision lattice QCD simulation
 is built on the assumption that one can
do a simulation in which the coupling is perturbative at the shortest available distances,
and the system will become nonperturbative at the longest ones.
This can can be observed in (for example) a heavy quark potential behaving as 
$V(r) \sim 1/r$ at small $r$ and $\sigma r$ for large $r$.
The ever bigger lattices used in state of the art simulations are present both to push the simulation
volume (measured in centimeters${}^4$)  ever larger, while simultaneously shrinking
 the lattice spacing ever smaller. ``Eliminating lattice artifacts''
 is a coded phrase for writing down a theory at the cutoff scale
which is continuum QCD, up to small and calculable corrections, which implies that it
 is weakly interacting there.

The slow running of the coupling for this model means that  on any lattice size accessible
for numerical simulations, the zero bare quark mass system is conformal for all practical
 purposes.
What does control the correlation length (inverse of a mass) is the fermion mass.
The scaling of the correlation length with fermion mass is given by a critical exponent
$y_m$ which is related to the anomalous dimension $\gamma_m$ of the  operator $\bar\psi\psi$,
defined through the running
 of the fermion mass
\bee
\mu \frac{\partial m(\mu)}{\partial \mu} = -\gamma_m(g^2) m(\mu),
\ee
The quantity $\gamma_m$ is the interesting parameter for technicolor dynamics.
Technicolor scenarios, conventional or not,
are said to prefer to have $\gamma_m\sim 1$ to simultaneously generate phenomenologically
 interesting fermion masses
while suppressing flavor changing neutral currents. 
(Compare the recent discussion in Ref.~\cite{Chivukula:2010tn}.)
Our theory is not very desirable from a phenomenological point of view:
 two studies\cite{DeGrand:2009hu,DSS} show that $\gamma_m$ is small,
about $0.35$ at the parameter values of the simulation.

Finally, to complete the list of undesirable features of the system, as
 one tunes the bare gauge coupling larger and larger, it  undergoes a first order
 transition
into a confining phase where the axial Ward identity
 quark mass never vanishes. It is not known whether this transition 
is a lattice artifact, or not. For the purpose of this paper, the resolution of this question is
only of indirect importance. 
At its bare coupling value, the lattice system has a slowly running coupling constant,
no chiral symmetry breaking, and no confinement. Let us just regard it as a template for
some exotic new physics scenario, and see what it produces for an electroweak observable:
what is  $\Pi^{LR}_T(q^2)$?

Readers should note: most lattice calculations are about numbers and precision.
This calculation has neither. Rather, it is a qualitative study of a theory which does not
resemble QCD. To
try to discover what it does resemble, I (mostly) compute $\Pi^T_{LR}$ rather than the $S$ parameter,
because at the bare couplings where I did the simulation, there is no $S$ parameter.
The true electroweak observable $S$ involves subtracting the contribution
 of the Higgs field -- or of the particles which 
replace the Higgs field -- from $\Pi_T^{LR}$. There is no evidence of
 Higgs - like dynamics in the weak coupling phase of this theory.

The realization of Eq.~\ref{eq:continPi}  in a lattice simulation involves
 a long stream of annoying technical problems which must be overcome. Fortunately,
 there are already two lattice calculations of the $S$ parameter in QCD,
 Refs.~\cite{Shintani:2008qe} and \cite{Boyle:2009xi}, plus earlier lattice work
on current-current correlators \cite{Gockeler:2000kj,Blum:2002ii}  which provide pretty 
explicit directions
to follow. The problems to be addressed are:
\begin{itemize}
\item Lattice artifacts in Eq.~\ref{eq:continPi}: in a lattice calculation
\bee
\Pi_{\mu\nu}(q) = P^T_{\mu\nu}(q) \Pi^T(q) + P^L_{\mu\nu}(q) \Pi_L(q) + \dots
\label{eq:dirt}
\ee
where $P_T(q)$ and $P_L(q)$ are lattice analogs of the transverse and longitudinal projectors and the dots represents
additional momentum-dependent terms, proportional to higher powers of products of $q$.
\item If the lattice currents are not conserved, there is an additional quadratic divergence
$\delta_{\mu\nu}Q/a^2$ in $\Pi_{\mu\nu}(q)$.
\item If the lattice currents are not local, there are additional contact terms in $\Pi_{\mu\nu}(q)$.
\item There is a lattice - to - continuum regularization factor for each current.
If it is different for the vector and axial vector currents, the analysis of a quantity 
depending on their difference becomes more fraught.
\end{itemize}

If we are only interested in the difference of vector and axial vector currents,
the use of valence fermions which encode exact chiral symmetry (overlap or domain wall fermions)
 alleviates the second, third, and fourth of these problems.

The next section describes technical details of the simulations. Readers uninterested in
 them should jump to the following two sections, for qualitative comparisons of
 $\Pi_T^{LR}$ from ordinary QCD and the walking theory.

\section{Numerical techniques and background\label{sec:two}}

I will make qualitative comparisons of lattice data from simulations of ordinary QCD
and of $SU(3)$ gauge theory coupled to two flavors of fermions in the symmetric (sextet)
representation. In all these studies, the
valence Dirac operator
 is taken to be the overlap
operator \cite{Neuberger:1997fp,Neuberger:1998my}.
 Details of the particular implementation of the action are described in
 Refs.~\cite{DeGrand:2000tf,DeGrand:2004nq,DeGrand:2006ws,DeGrand:2007tm,DeGrand:2006nv};
suffice it to say that the massless overlap operator is defined as 
$D=R_0(1+ d(-R_0)/\sqrt{d^\dagger(-R_0)d(-R_0)})$
where $d(m)=d+m$ for some lattice approximation (``kernel'') $d$ to the massless
continuum Dirac operator. 
The only new ingredient is the application to symmetric-representation fermions,
already described in Ref.~\cite{DeGrand:2009hu}.
Eigenvalues of the squared Hermitian Dirac operator $D^\dagger D$ are computed
 using the ``Primme'' package of
McCombs and Stathopoulos\cite{primme} and are used to precondition the calculation of propagators.

The lattice analog of Eq.~\ref{eq:continPi}
 is computed using the difference of improved currents, that is, the vector current is
\bee V_\mu^{12}= \bar q_1 \gamma_\mu(1-\frac{aD}{2R_0}) q_2 . \ee
The axial current is identical, apart from the substitution of
 $\gamma_\mu \gamma_5$ for $\gamma_\mu$.
These currents are not conserved, and so the correlator of each current is quadratically divergent.
However, because  they are related by a Ward identity, the quadratic divergence
cancels in the vector-axial difference. In addition, 
 both currents have the same lattice-to-continuum regulator
 renormalization factor (Z-factor). The currents are local enough that $\Pi_{\mu\nu}$ 
does not have contact terms.

 In practice, the correlator is computed
 using point currents and the
``shifted'' propagator
\bee
\hat D^{-1}(m_q) = \frac{1}{ 1-m_q/(2R_0)}(D^{-1}(m_q) - \frac{1}{2R_0}) .
\label{eq:SUBPROP}
\ee
In free field theory and to all orders in perturbation theory $\Pi^{LR}=0$
at $m_q=0$, basically because
$\{\gamma_5,\hat D^{-1}\}=0$.

This gives a sensible and useful $\Pi_{\mu\nu}^{LR}$
 -- a correlator of local currents with a common $Z$ factor, and with the decomposition of
Eq.~\ref{eq:continPi} for small momenta.
 I now have to deal with the lattice artifacts in the decomposition of
Eq.~\ref{eq:dirt}. Following Ref.~\cite{Shintani:2008qe}, I observe that they are small. This is 
quantified
via the observable
\bee
\Delta_J(q)= \sum_{\mu\nu} \bar q_\mu \bar q_\nu (\frac{1}{\bar q^2} - \frac{\bar q_\nu}{\sum_\lambda \bar q_\lambda^3} ) \Pi_{\mu\nu}^J(q)
\ee
In a slight variation on the method of Ref~\cite{Shintani:2008qe},
 I define the appropriate variable as
$\bar q_\mu=(2/a)\sin q_\mu a/2)$. (Of course,
 $q_\mu= (2\pi/L)n_\mu$ for integer $n_\mu$ if the $\hat \mu$ direction
exhibits periodic boundary conditions and its length is $L$.)
A $\Pi_{\mu\nu}$ which is a superposition of pure longitudinal and transverse terms (the continuum
decomposition of Eq.~\ref{eq:continPi} with $q$'s replaced by $\bar q$'s) will give $\Delta=0$.
When I come to analysis I will show that $\Delta$ is very small compared to $\Pi_T^{LR}$.

Next, I have to perform the decomposition of $\Pi_{\mu\nu}$ into $\Pi_L$ and $\Pi_T$.
 An easy way to do this is
to assume that
\bee
\Pi_{\mu\nu}(q)= P^T_{\mu\nu}(q)\Pi_T(q)+P^L_{\mu\nu}(q)\Pi_L(q)
\ee
where 
\bee
P^T_{\mu\nu}(q)= \bar q^2 \delta_{\mu\nu} - \bar q_\mu \bar q_\nu
\ee
and 
\bee
P^L_{\mu\nu}(q)=\bar q_\mu \bar q_\nu
\ee
are the transverse and longitudinal projectors.
 Next I  form the chi-squared function, individual momentum mode
by momentum mode, taking the sixteen  $(\mu\nu)$ correlators as the quantities to be fit,
\bee
\chi^2(q)=\sum_{\mu\nu}(\Pi_{\mu\nu}(q)-P^T_{\mu\nu}(q)\Pi_T(q)-P^L_{\mu\nu}(q)\Pi_L(q))^2,
\ee
and treating $\Pi_T$ and $\Pi_L$ as fit parameters. Because $P^T$ and $P^L$ are
projectors, $\Tr P^T P^L=0$,  this minimization reduces to simple definitions of weighted averages,
\beea
\Pi_T(q)  &=& \frac{\sum_{\mu\nu}P^T_{\mu\nu}(q)\Pi_{\mu\nu}(q)}{3(\bar q^2)^2} \nonumber \\
\Pi_L(q) &=& \frac{\sum_{\mu\nu}P^L_{\mu\nu}(q)\Pi_{\mu\nu}(q)}{(\bar q^2)^2}  . \nonumber \\
\label{eq:fit}
\eea

Of course, all lattice data from the same set of configurations are highly correlated.
The uncertainties in the pictures which follow are computed
 by folding  Eq.~\ref{eq:fit} into a single elimination jackknife:
I delete a lattice from my data set, compute $\Pi_T$, $\Pi_L$, and $\Delta$,
 and then present a jackknife 
average and uncertainty.

\section{Illustrations from QCD}

The remainder of the paper is a qualitative comparison of $\Pi_T^{LR}$ from ordinary QCD and from
sextet QCD. I begin with ordinary QCD.

I have two data sets. The first is a quenched set
of 20 lattices computed using the overlap operator for valence quarks
on a background of Wilson gauge action configurations at a gauge coupling $\beta=5.9$. The lattice
size is $16^4$ sites. The lattice spacing  is about 0.13 fm from the rho mass or
 0.11 fm from the Sommer parameter.
This data set can be compared to an extensively-analyzed companion  used in a calculation of
$B_K$, the kaon B-parameter \cite{DeGrand:2003in}. (The parameters of the kernel action used in this 
study differ by
about one per cent from those of the $B_K$ project, but that is not going to matter 
given the qualitative nature of my presentation.)

The second data set is a smaller volume ($12^4$ sites) set of simulations with $N_f=2$ flavors
of dynamical overlap fermions. The data set had three quark masses, $am_q=0.03$, 0.05, 0.10,
at a lattice spacing of roughly 0.14 fm from the Sommer parameter. These lattices have been used
in several small projects \cite{DeGrand:2007tx} to date.
They are not really big enough for reliable spectroscopy measurements.

I cannot extract an $S$ parameter from these results (though I tried). The reason is that to get
$S$ from $\Pi_T^{LR}$ requires doing a fit of the data
to a chiral perturbation theory calculation and removing the single pseudoscalar propagator 
contribution. $S$ is then related to one of the Gasser-Leutwyler $O(p^4)$ coefficients.
Doing this requires going to very low $q^2$, so that chiral perturbation theory is in its
domain of applicability. This in turn requires with present techniques
making the lattice volume large. 
(For example, Ref.~\cite{Shintani:2008qe} used a $16^3\times 32$ lattice.)
My $16^4$ lattice is large but it is a quenched data set and I am unaware of the appropriate
quenched chiral perturbation theory calculation. And of course, the quenched approximation is obsolete
as a potential high-precision venue. For this study, I do not need $S$.

\begin{figure}
\begin{center}
\includegraphics[width=0.6\textwidth,clip]{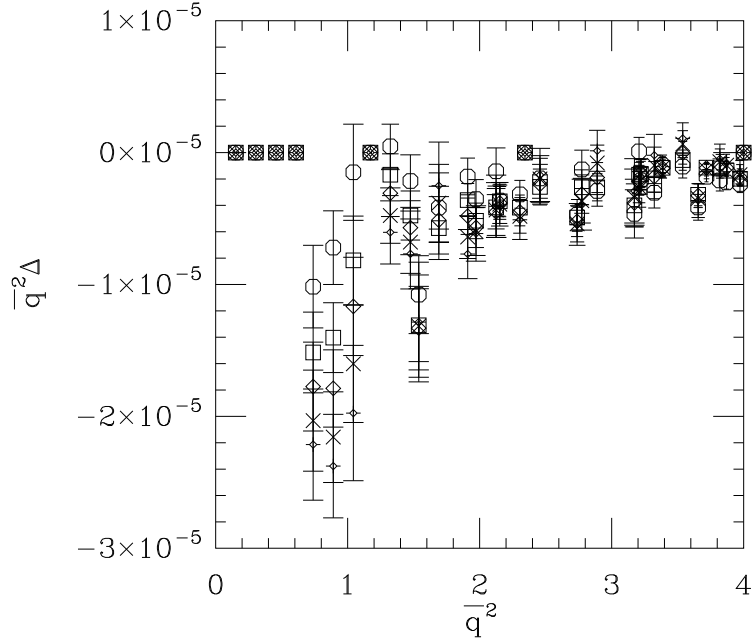}
\end{center}
\caption{$\bar q^2\Delta^{V-A}$ vs $\bar q^2$  vs $\bar q^2$ from quenched
 overlap fermions  at $\beta=5.9$.
Valence masses are (octagons),
 $am_q=0.10$ (squares), $am_q=0.05$ (diamonds),
 $am_q=0.035$. (crosses) $am_q=0.025$,
and (fancy diamonds) $am_q=0.02$.
\label{fig:jlqcdQ}}
\end{figure}

\begin{figure}
\begin{center}
\includegraphics[width=0.6\textwidth,clip]{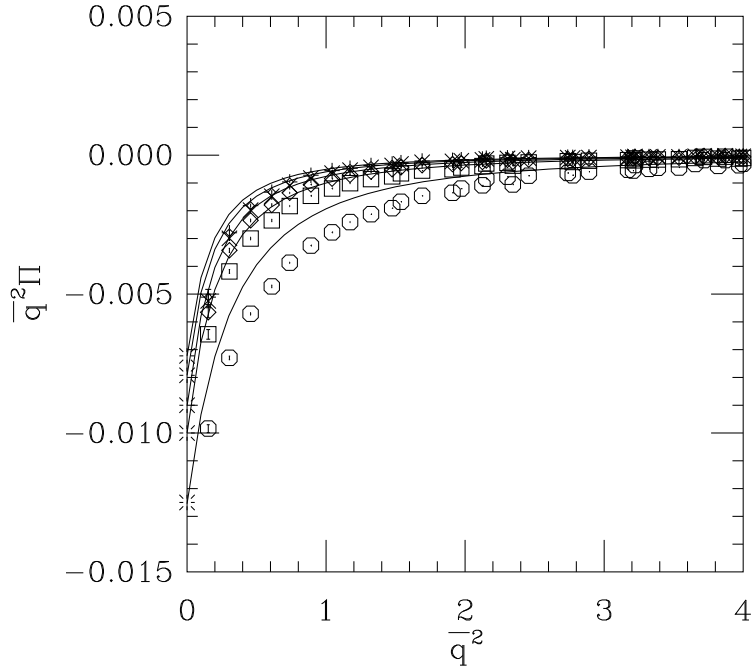}
\end{center}
\caption{$\bar q^2\Pi_T^{V-A}$ vs $\bar q^2$ from quenched overlap fermions  at $\beta=5.9$,
As in Fig.~\protect{\ref{fig:jlqcdQ}}, valence masses are shown by 
octagons for $am_q=0.10$,
squares for $am_q=0.05$,
diamonds for $am_q=0.035$,
crosses for $am_q=0.025$,
and fancy diamonds for $am_q=0.02$.
 The lines are the model parameterizations.
\label{fig:transQ}}
\end{figure}

Figs.~\ref{fig:jlqcdQ} and \ref{fig:transQ} show the $\Delta$ parameter and $\Pi_T^{LR}$ from the 
quenched data set. In keeping with past practice \cite{Shintani:2008qe,Boyle:2009xi},
 I have plotted $\bar q^2 \Pi_T^{LR}$.
$\Delta$ seems satisfactorily small compared to $ \Pi_T^{LR}$.
   Unsurprisingly, $\Pi_T^{LR}$ qualitatively resembles other 
published results.

\begin{figure}
\begin{center}
\includegraphics[width=0.6\textwidth,clip]{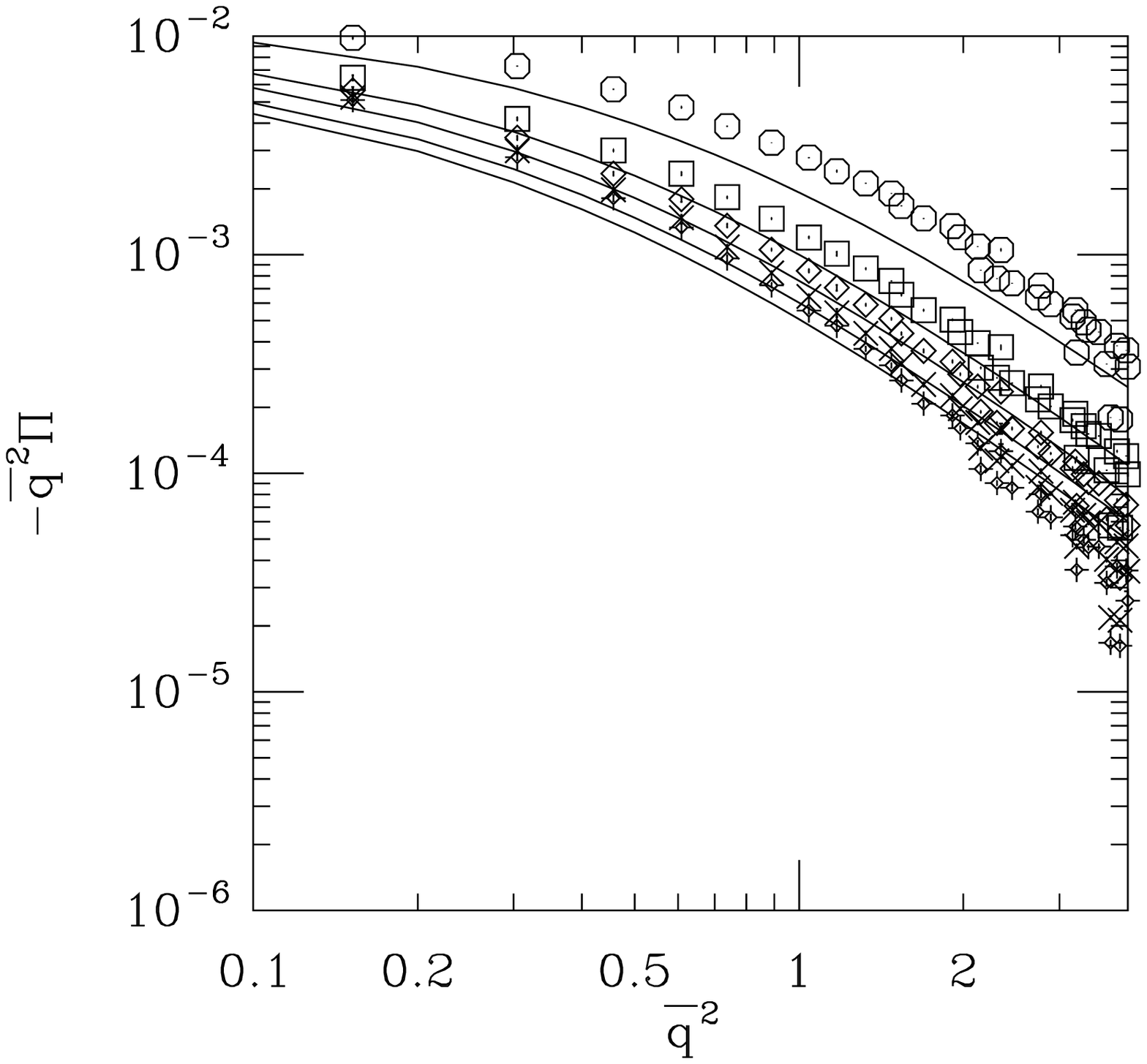}
\end{center}
\caption{As in Fig.~\protect{\ref{fig:transQ}}, but a logarithmic scale to
 show the approach to Weinberg sum rules.
\label{fig:transLQ}}
\end{figure}

The superimposed curves in Fig.~\ref{fig:transQ} show phenomenological parameterizations
 of the data, where
$\Pi^{LR}_T$ is saturated with a set of resonances. Introducing two quark flavor labels
to characterize the currents, the pseudoscalar decay constant $f_\pi$ is defined through
\bee
\langle 0| \bar u \gamma_0 \gamma_5 d |\pi\rangle = m_\pi f_\pi
\ee
(so $f_\pi\sim 132$ MeV)
while the vector meson decay constant of state $V$ is defined as
\bee
\langle 0| \bar u \gamma_i d  | V\rangle = m_V^2 f_V \epsilon_i
\ee
and the axial vector meson decay constant of state $A$ is
\bee
\langle 0|  \bar u \gamma_i \gamma_5 d  |A \rangle = m_A^2 f_A \epsilon_i.
\ee
$\epsilon_i$ is a unit polarization vector.
The transverse current correlator is then
\bee
\Pi^{LR}_T(q^2) = \sum_V\frac{f_V^2M_V^2}{q^2+M_V^2} - \sum_A \frac{f_A^2M_A^2}{q^2+M_A^2} - \frac{f_\pi^2}{q^2}.
\label{eq:SR}
\ee
In a theory with spontaneous chiral symmetry breaking, the couplings are constrained by the
first 
\bee
\sum_V f_V^2M_V^2 - \sum_A f_A^2M_A^2 -f_\pi^2 =0
\label{eq:W1}
\ee
and second
\bee
\sum_V f_V^2M_V^4 - \sum_A f_A^2 M_A^4 =0
\label{eq:W2}
\ee
 Weinberg sum rules \cite{Weinberg:1967kj}. When the two sum rules are satisfied, $\Pi^{LR}_T(q^2)$
falls off asymptotically in $q$ as $1/q^6$.

A common phenomenological approach is to saturate the correlator by the lowest states in the vector
and axial channels, the pion, rho and $a_1$ meson\cite{MHA}. This involves fixing five parameters,
the three couplings $f_\pi$, $f_\rho$, and $f_{a_1}$ and two masses $m_\rho$ and $m_{a_1}$.
In principle, all of these can be measured in lattice simulations.
 In practice, $f_{a_1}$ and  $m_{a_1}$
involve difficult measurements because the signal to noise ratio in a mesonic correlator
(whose mass is $M$) scales
like $\exp((M-2m_\pi)t)$. 

I attempted to extract all five quantities from spectroscopic fits to the quenched data.
I could do this only for the heavier masses  in the data set.
Results are shown in Table \ref{tab:Q}.
Using them in the correlator
 does not reproduce either the data or the Weinberg sum rules. (Perhaps this result is neither
surprising nor controversial.) However, I do want to compare the QCD results to
those from the walking theory.
Accordingly, one approach to parameterizing the data consists of taking the values
 of $f_\pi$, $m_\rho$,
and $f_\rho$ from fits to lattice data, and determining $f_{a_1}$ and $m_{a_1}$ by forcing
a solution to the to Weinberg sum rules.

The result of this calculation is shown as the curves in Fig.~\ref{fig:transQ}.
In this figure, the bursts at $\bar q=0$ are just the values of $f_\pi^2$.
Fig.~\ref{fig:transLQ} shows $\bar q^2\Pi_T^{V-A}$ on a log-log scale to expose the approach to
scaling at large $\bar q$. The data qualitatively resembles the analytic parameterization.
The Weinberg sum rules are statements about the asymptotic behavior of the current correlators.
Lattice QCD correlators presumably attempt to encode these statements until they are overwhelmed by
 lattice artifacts.

\begin{figure}
\begin{center}
\includegraphics[width=0.6\textwidth,clip]{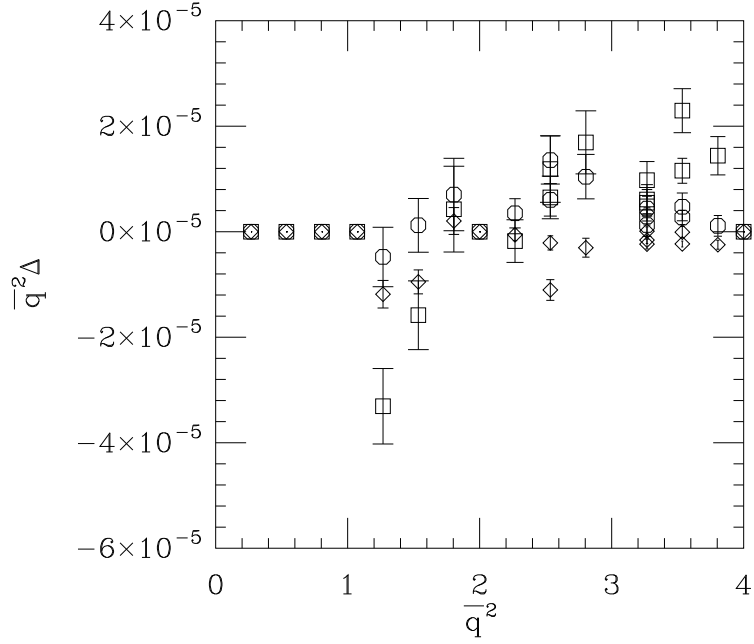}
\end{center}
\caption{$\bar q^2\Delta^{V-A}$ vs $\bar q^2$  vs $\bar q^2$  from dynamical $N_f=2$ simulations.
Valence masses are (octagons) $am_q=0.10$,
 (squares) $am_q=0.05$,
 (diamonds) $am_q=0.03$.
\label{fig:jlqcdNF2}}
\end{figure}

\begin{figure}
\begin{center}
\includegraphics[width=0.6\textwidth,clip]{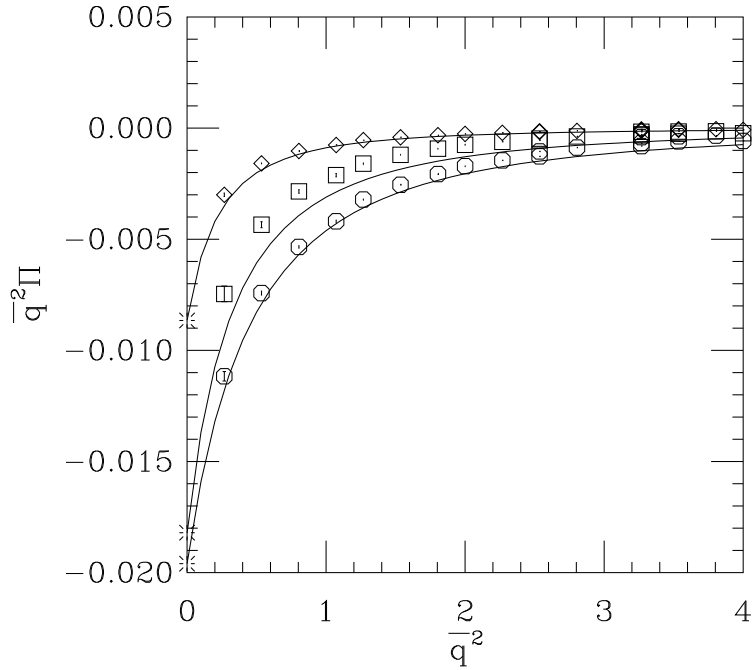}
\end{center}
\caption{$\bar q^2\Pi_T^{V-A}$ vs $\bar q^2$ from dynamical $N_f=2$ simulations.
Valence masses are (octagons) $am_q=0.10$,
 (squares) $am_q=0.05$,
 (diamonds) $am_q=0.03$. The lines are the model parameterizations.
\label{fig:transNF2}}
\end{figure}

\begin{figure}
\begin{center}
\includegraphics[width=0.6\textwidth,clip]{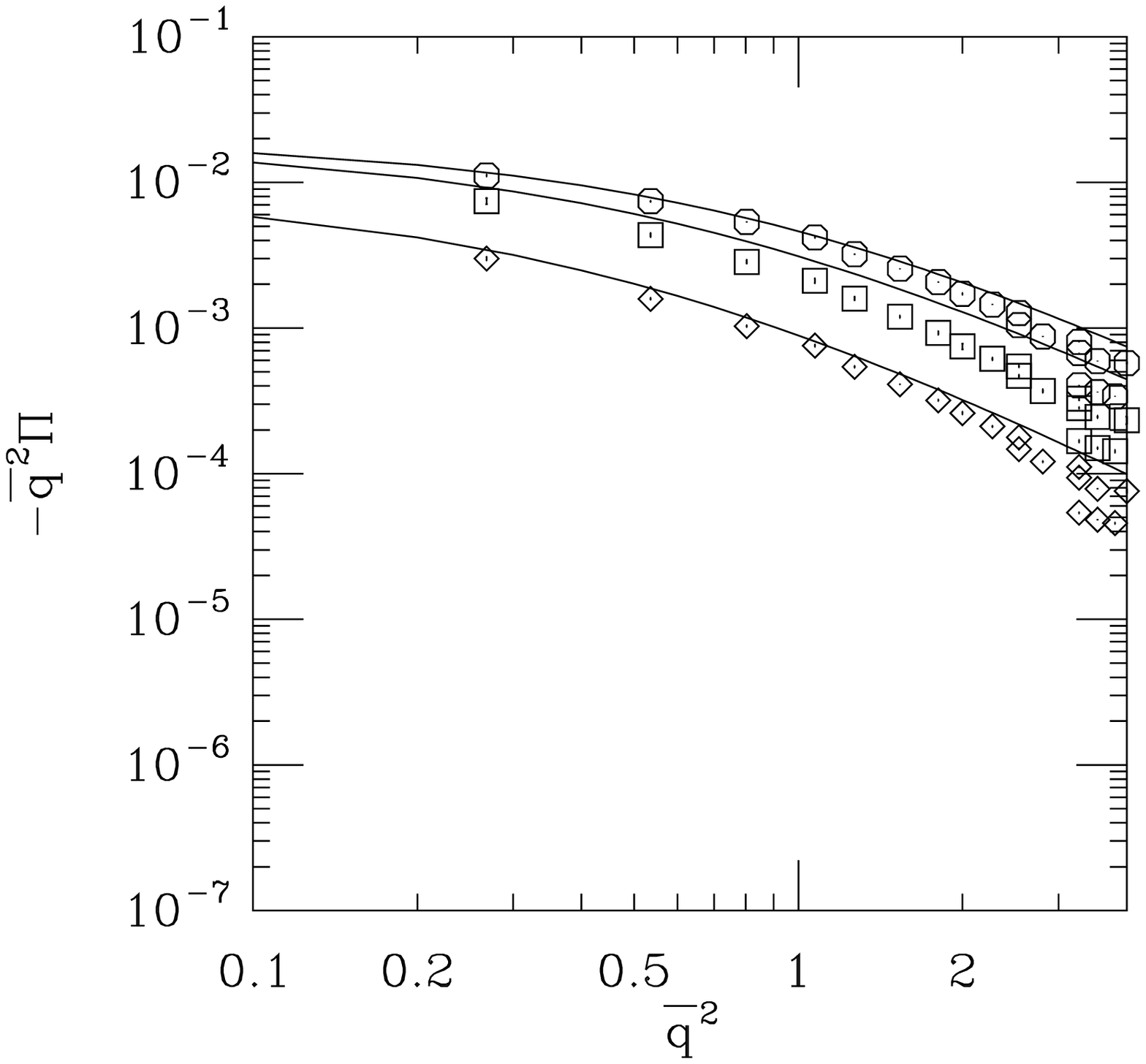}
\end{center}
\caption{As in Fig.~\protect{\ref{fig:transNF2}}, but a log scale to show the
 approach to Weinberg sum rules.
\label{fig:transLNF2}}
\end{figure}

Next, we turn to figures from the
 $12^4$ volume dynamical overlap:
Fig.~\ref{fig:jlqcdNF2} shows the $\Delta$ observable.
Fig.~\ref{fig:transNF2} shows $\bar q^2\Pi_T^{V-A}$ vs $\bar q^2$. Again, the superimposed curves
are from Eq.~\ref{eq:SR}, taking fit values of $f_\pi$, $f_\rho$ and $m_\rho$, 
and determining $f_{a_1}$
and $m_{a_1}$ by saturating the two Weinberg sum rules.
Fig.~\ref{fig:transLNF2} reproduces Fig.~\ref{fig:transNF2} on a log-log scale to show the
Weinberg sum
rules at work. As a qualitative parameterization of the data, this procedure works well.

\begin{table}
\begin{tabular}{c|ccccc|cccc}
\hline
$m_q$ & $f_\pi$     & $m_\rho$ & $f_\rho$ & $m_{a_1}$ & $f_{a_1}$ & fit  $m_{a_1}$ & fit $f_{a_1}$ &
$\overline L_{10}\times 10^3$ & $a^2\Delta m_\pi^2\times 10^3$ \\
\hline
0.020   & 0.085(1)  & 0.55(2)  & 0.282(5) & & &0.657(15) & 0.197(3) &5.1(6) & 0.62(4) \\
0.025   & 0.089(1)  & 0.56(2)  & 0.277(4) & &  &0.684(14) &0.185(9) &5.3(6) & 0.66(5)\\
0.035   & 0.095(1)  & 0.57(1)  & 0.271(3)  & &  &0.723(8) &0.168(6) &5.6(7) &0.71(3) \\
0.050   & 0.100(1)  & 0.60(1)  & 0.261(2) & 0.93(5) & 0.182(3) &0.780(8) &0.154(5) &5.5(8) & 0.81(3) \\
0.100   & 0.1168(5)  & 0.689(4)  & 0.235(2) &1.00(3) & 0.170(2) &0.995(10) &0.113(4) & 5.3(6) & 1.16(2) \\
\end{tabular}
\caption{\label{tab:Q} Measured lattice parameters (in lattice units; everything is scaled by
a factor of the lattice spacing $a$ and the overall $Z$ factor is left out)
 from quenched $16^4$ simulations, used in constructing
the parameterization of the transverse correlator. The last four columns are fit results from forcing the
spectral function to obey the first and second Weinberg sum rules, saturated by the rho and $a_1$ mesons.
 }
\end{table}

\begin{table}
\begin{tabular}{c|ccc|cccc}
\hline
$m_q$ & $f_\pi$     & $m_\rho$ & $f_\rho$  & fit  $m_{a_1}$ & fit $f_{a_1}$ &
$\overline L_{10}\times 10^3$ & $a^2 \Delta m_\pi^2\times 10^3$   \\
\hline
0.03 & 0.092(3)  & 0.57(2)  &  0.22(2)  & 0.84(10) &0.10(3) & 4.8(4) & 0.81(12) \\
0.05 & 0.135(5)  & 0.64(2)  &  0.25(2)  & 1.19(26)  & 0.07(4) & 7.1(7) & 1.24(23) \\
0.10 & 0.138(6) & 0.80(1)  &  0.23(2)   & 1.20(15) & 0.10(3) &5.3(5) &1.64(18) \\
\end{tabular}
\caption{\label{tab:nf2} Measured lattice parameters from dynamical $12^4$ simulations, used in
 constructing
the parameterization of the transverse correlator.  The last four columns are fit results from forcing the
spectral function to obey the first and second Weinberg sum rules, saturated by the rho and $a_1$ mesons.
 }
\end{table}

\begin{figure}
\begin{center}
\includegraphics[width=0.9\textwidth,clip]{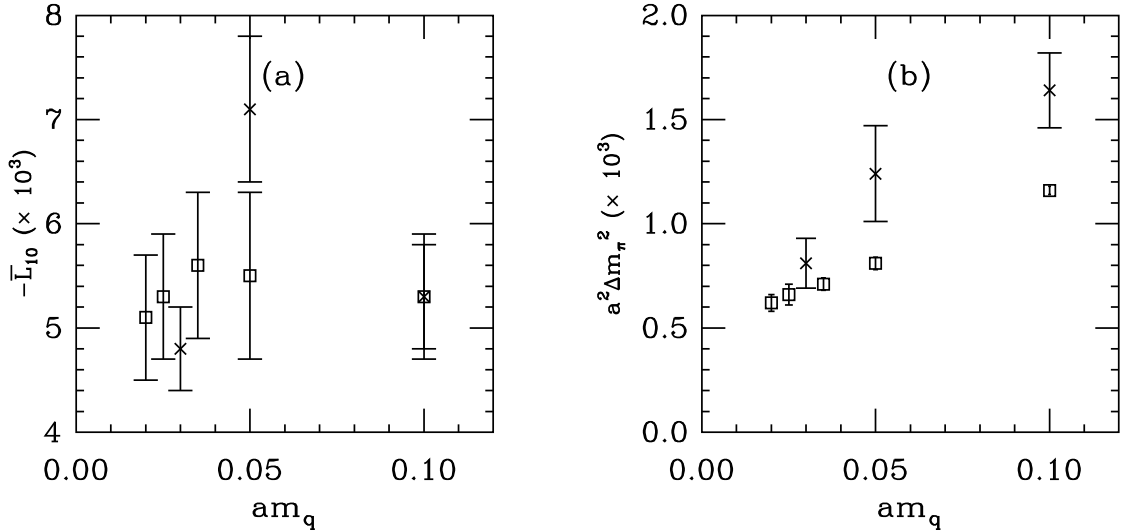}
\end{center}
\caption{
Panel (a) shows lattice data for $\overline L_{10}$, and panel (b), for $\Delta m_\pi^2$
from the analysis described in the text. Squares show quenched data and crosses are for $N_f=2$ simulations.
\label{fig:dgmly}}
\end{figure}

I cannot resist using the pseudoscalar and vector decay constants, plus the vector meson mass,
to compute the Gasser-Leutwyler parameter $\overline L_{10}$ and the
 charged - neutral pion mass difference $\Delta m_\pi^2$.
via the Das, Guralnik, Mathur, Low, Young sum rule \cite{Das:1967it}.
 In the ``lowest mass dominance'' approximation to the spectral
function, $\overline L_{10}= (f_\rho^2 - f_{a_1}^2)/8 = f_\rho^2(R^2-1)$ and
\bee
\Delta m_\pi^2 =
 \frac{3\alpha}{4\pi} \frac{1}{(\frac{1}{m_\rho^2} - \frac{1}{m_{a_1}^2})}\log \frac{m_{a_1}^2}{m_\rho^2}
 = \frac{3\alpha}{4\pi} \frac{m_\rho^2}{R-1}\log R
\ee
where $R= 1- (f_\pi/(m_\rho f_\rho))^2$ is given to remind the reader, what are
 the real independent variables. The results are shown in the two tables and in Fig.~\ref{fig:dgmly}.
A fit to a linear dependence on the quark mass gives $\overline L_{10}= 5.1(6)\times 10^{-3}$
and $5.3(5) \times 10^{-3}$ for the dynamical and quenched data sets. This is in 
good agreement with large scale simulation results of Refs.~\cite{Shintani:2008qe,Boyle:2009xi}
and with phenomenological estimates \cite{Bijnens:1994qh}.

Fitting the dynamical  data set's $\Delta m_\pi^2$ in the same way gives
 $a\Delta m_\pi^2 = 0.49(18)\times 10^{-3}$ or (with $a=0.14$ fm) 980(360) MeV${}^2$.
 The quenched data has 
 $a\Delta m_\pi^2 = 0.48(3)\times 10^{-3}$ for 1540(100) MeV${}^2$ or 1100(70) MeV${}^2$,
taking $a=0.11$ or 0.13 fm. These results agree nicely with experiment, 1261 MeV${}^2$.
Readers should note that while these are lattice results, they are indirect and depend on the
additional assumption that the spectral functions can be saturated by three input
 parameters plus the two
Weinberg sum rules.

The summary of this section is that both quenched and dynamical lattice QCD possesses
 a $q^2\Pi_T^{V-A}(q^2)$ which at
 small $q$ remains finite as the quark mass vanishes (it reduces to $-f_\pi^2$).
At  large $q$ it vanishes
roughly in accordance with the Weinberg sum rules. A simple three-parameter
 function using information from the lowest
excitations in the axial and vector channels seems to reproduce the data reasonably well
and produce phenomenologically sensible results.

\section{$\Pi^{LR}$ from sextet fermions}

I repeat the calculation of $q^2\Pi_T^{V-A}(q^2)$ using sextet representation overlap
fermions on a background of 
dynamical simulations
of $N_f=2$ flavors of clover fermions in the sextet representation and $SU(3)$ gauge fields.
The dynamical simulations were performed using the Wilson gauge action at a coupling $\beta=5.2$
 and a hopping parameter
$\kappa=0.1285$. This corresponds to an AWI (axial Ward identity) quark mass of $am_q^s=0.044$.
I used five valence masses, $am_q=0.100$, 0.050, 0.035, 0.025 and 0.020. 
The lattice volume is $16^4$ sites. At this set of simulation parameters,
 the system seems to be deconfined (the string tension is immeasurably small) and
chiral symmetry is restored (observed through the parity doubling of states).

\begin{figure}
\begin{center}
\includegraphics[width=0.6\textwidth,clip]{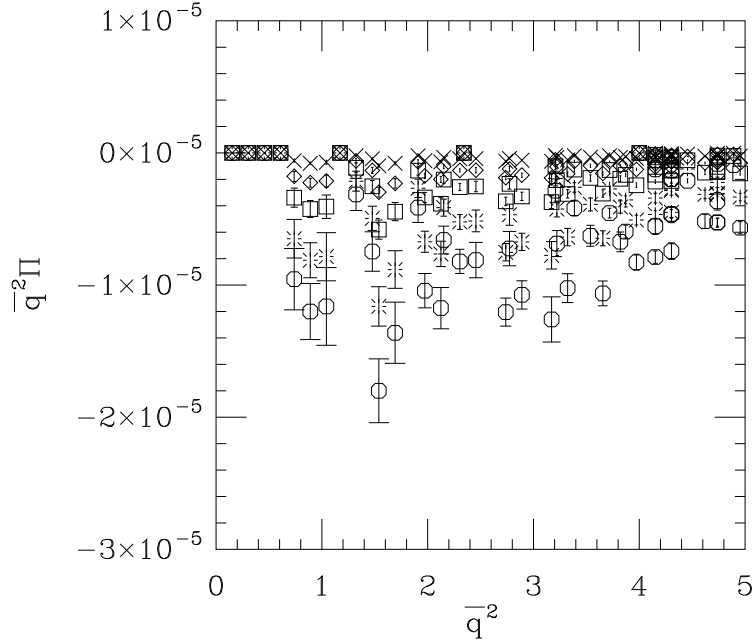}
\end{center}
\caption{$\hat q^2\Delta^{V-A}$ vs $\hat q^2$ from valence sextet overlap fermions on background
sextet dynamical
clover simulations at $\beta=5.2$,
$\kappa=0.1285$.  Valence masses are (octagons) $am_q=0.10$, (bursts) $am_q=0.075$, (squares) $am_q=0.05$,
(diamonds) $am_q=0.035$. (crosses) $am_q=0.020$.
\label{fig:jlqcd}}
\end{figure}

\begin{figure}
\begin{center}
\includegraphics[width=0.6\textwidth,clip]{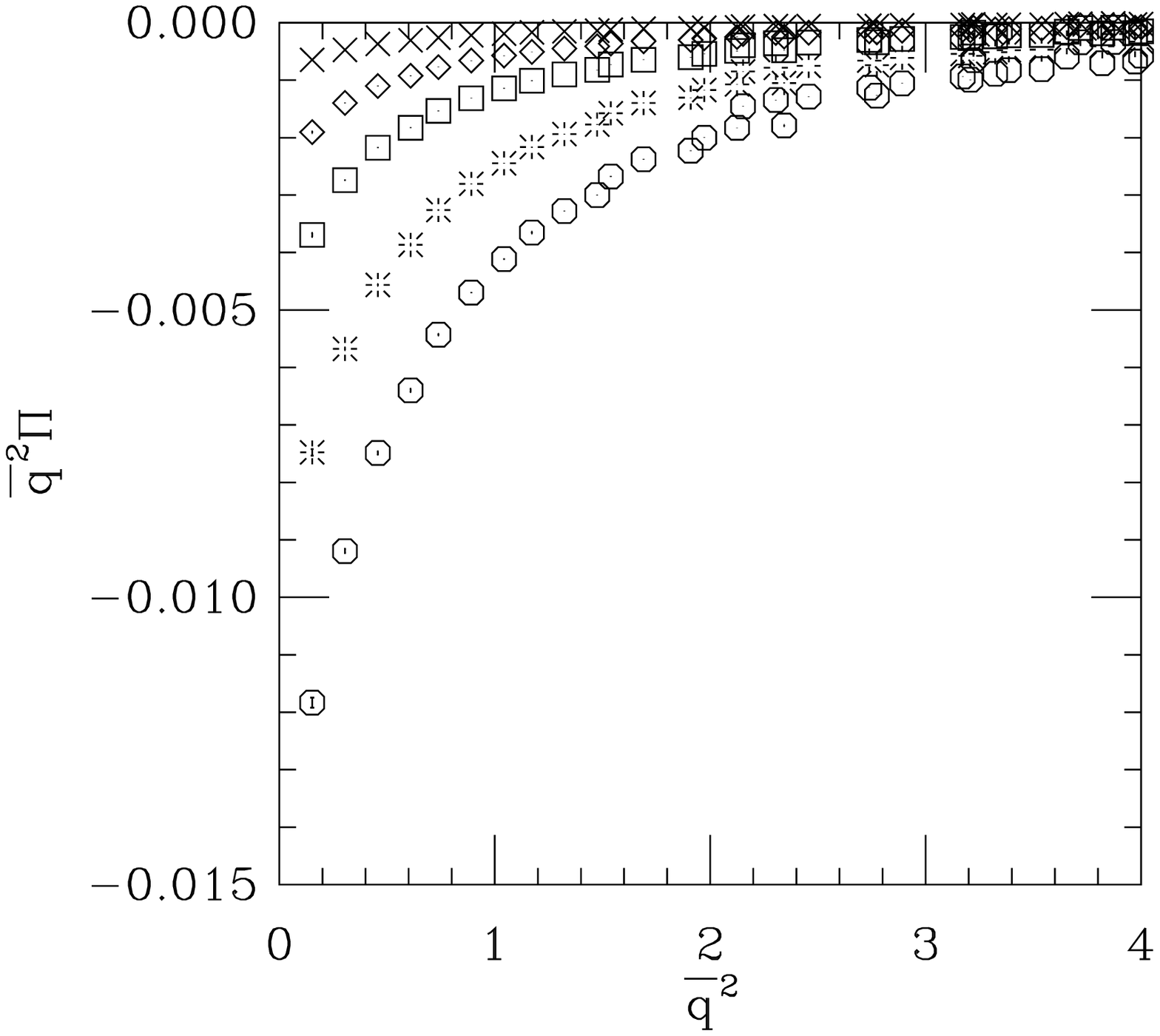}
\end{center}
\caption{$\bar q^2\Pi_T^{V-A}$ vs $\hat q^2$ from valence sextet overlap fermions on background
sextet  dynamical clover simulations at $\beta=5.2$,
$\kappa=0.1285$. Valence masses are (octagons) $am_q=0.10$, (bursts) $am_q=0.075$, (squares) $am_q=0.05$, (diamonds) $am_q=0.035$. (crosses) $am_q=0.020$.
\label{fig:trans}}
\end{figure}

\begin{figure}
\begin{center}
\includegraphics[width=0.6\textwidth,clip]{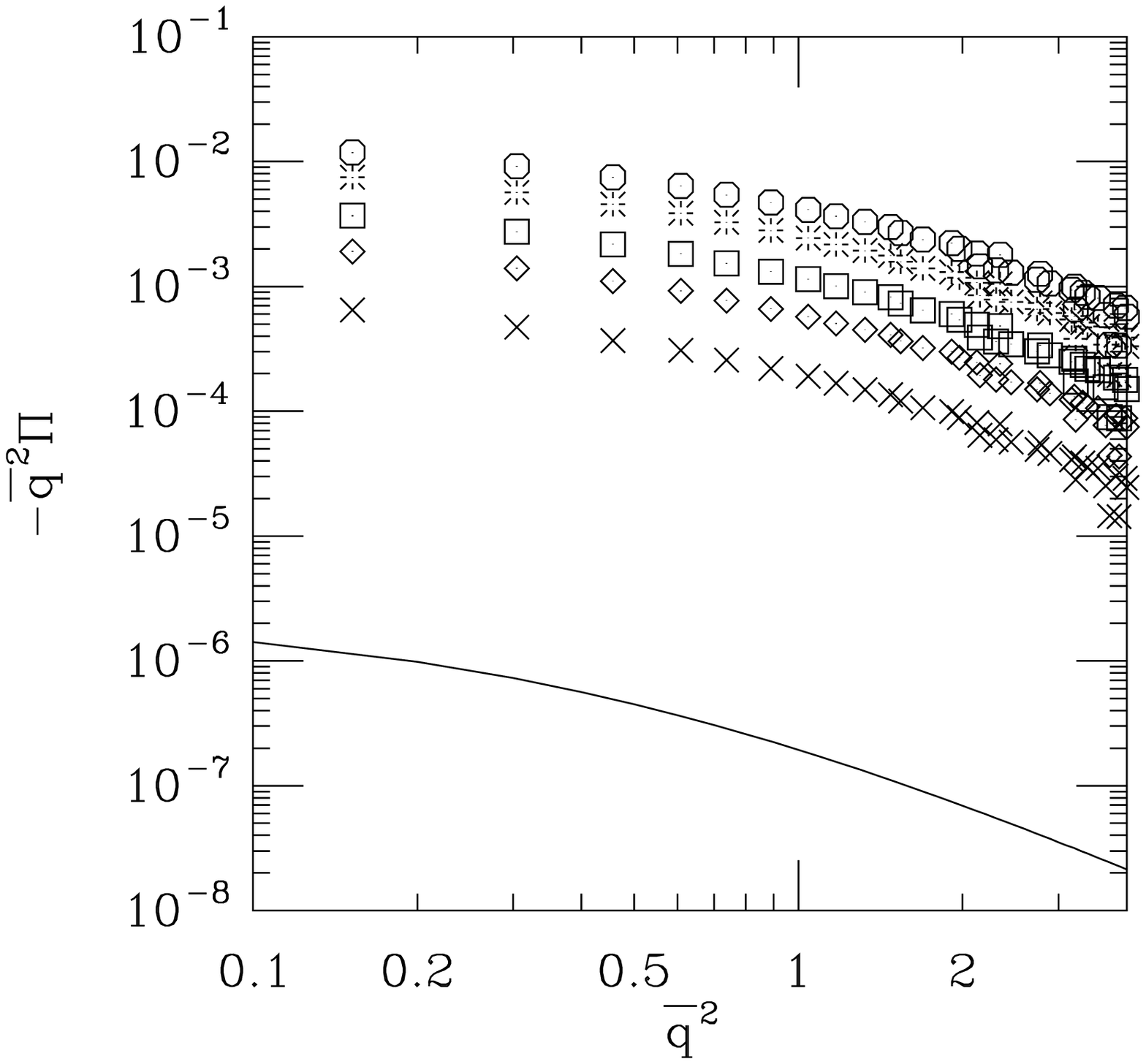}
\end{center}
\caption{As in Fig.~\protect{\ref{fig:trans}}, but a log-log scale.
The line shows an attempted  parameterization of the data, forcing the
 two Weinberg sum rules to be saturated
by the three lowest resonances.
\label{fig:transL}}
\end{figure}

Fig.~\ref{fig:jlqcd} shows the $\Delta$ observable. It seems to be acceptably small.
Fig.~\ref{fig:trans} shows $\bar q^2\Pi_T^{V-A}$ vs $\hat q^2$.
Fig.~\ref{fig:transL} shows $-\bar q^2\Pi_T^{V-A}$ vs $\hat q^2$, replotted with logarithmic axes.
These figures show very different behavior from the case of QCD.

Notice that in the large-$q^2$ limit, $\Pi^{LR}_T$ still falls to zero. This is consistent with the
Weinberg sum rules, and is expected for any theory with
an ultraviolet-attractive fixed point at $g=0$ 
\cite{Bernard:1975cd}.

Just as in the case of QCD,
it is possible to measure spectroscopy in the pseudoscalar, vector, and axial vector channels.
 It is also possible to measure
the three decay constants. In fact, it is easier to extract the axial vector meson properties here,
than it is in an ordinary QCD simulation. This is because the system is chirally restored and the pion is not
appreciably lighter than the other states. (The same effect occurs in the high temperature phase of QCD.)
 These parameters are shown in Table~\ref{tab:sextet}. The obvious
 feature they show is chiral symmetry restoration: note the near degeneracy of vector and axial vector
masses (called $m_\rho$ and $m_{a_1}$ in hadronic analogy), which becomes more pronounced as the fermion mass vanishes) and of the vector and axial vector couplings.

A quick check with a calculator shows that the three lowest states with their couplings 
do not saturate either Weinberg sum rule. So I play the same game that I did for the QCD data sets,
determining $m_{a_1}$ and $f_{a_1}$ by forcing the two sum rules to saturate.
The result of this exercise is shown in Fig.~\ref{fig:transL}  (for the $m_q=0.100$
data set, only). This fails completely.

I was somewhat surprised by this result, since lowest mass dominance is a venerable ingredient of strongly
coupled beyond - Standard Model phenomenology  (compare Ref.~\cite{Peskin:1991sw}, or for a recent review, 
Ref.~\cite{Contino:2010rs}), even sometimes
of conformal beyond - Standard Model phenomenology (compare Refs.~\cite{Appelquist:1998xf,Sannino:2010ca}.
However, note that I am looking at a system in which the breaking of (near) conformal symmetry
is by a small mass, and I am looking at $m^2/q^2$ small.
Often, the phenomenology is concerned with the other limit,  $m^2/q^2$ large.

\begin{table}
\begin{tabular}{c|ccccc}
\hline
$m_q$ & $f_\pi$     & $m_\rho$ & $f_\rho$ & $m_{a_1}$ & $f_{a_1}$ \\
\hline
0.020 & 0.00050(4) & 0.33(1) & 0.0046(2) & 0.38(1) & 0.0044(2) \\
0.035 & 0.00080(6)  & 0.37(1) & 0.0045(2)  & 0.42(2) & 0.0043(1)\\
0.050 & 0.00102(8)  & 0.37(1) & 0.0044(2) & 0. 47(2) & 0.0042(2) \\
0.075 & 0.00131(9) & 0.43(1) &0.0041(1)  & 0.54(2) & 0.0037(10)\\
0.100 & 0.00151(10) & 0.48(1) & 0.0037(1) & 0.61(2) & 0.0037(2) \\
\end{tabular}
\caption{\label{tab:sextet} Measured lattice parameters from valence overlap fermions
from sextet QCD simulations.
 }
\end{table}

\begin{figure}
\begin{center}
\includegraphics[width=0.6\textwidth,clip]{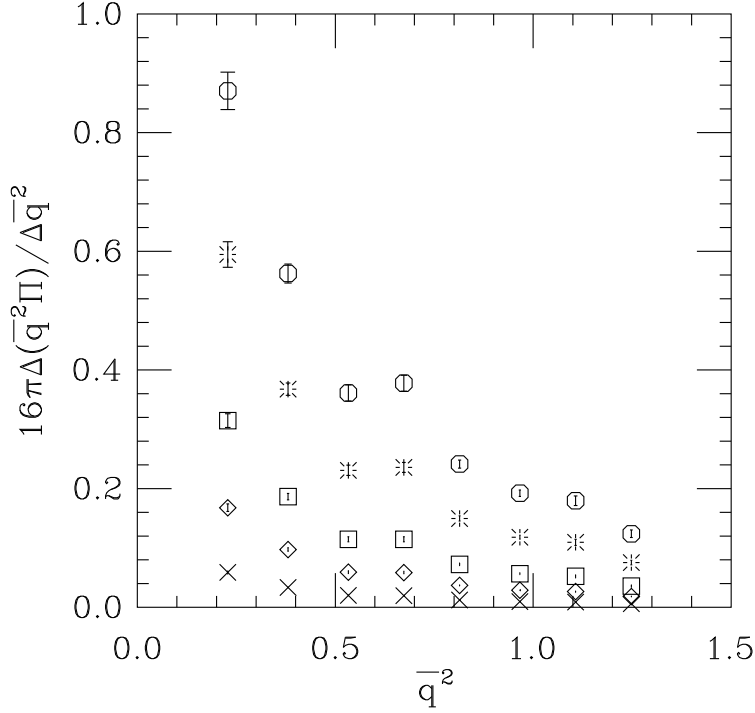}
\end{center}
\caption{Derivative of the sextet QCD correlator
$S_{eff}=16\pi\Delta(\overline q^2\Pi_T^{V-A})/\Delta \overline q^2$..
Valence masses are (octagons) $am_q=0.10$, (bursts) $am_q=0.075$, (squares) $am_q=0.05$,
(diamonds) $am_q=0.035$. (crosses) $am_q=0.020$. 
\label{fig:xtransl}}
\end{figure}

 I convert
$\Pi_T^{V-A}$ into an ``effective'' S-parameter by numerically differentiating it,
\bee
S_{eff}(q^2) = 16\pi\frac{\Delta(\overline q^2\Pi_T^{V-A})}{\Delta \overline q^2}
\ee
(For the resonance model, Eq.~\ref{eq:SR}, this would give $S_{eff}=16\pi(f_V^2-f_A^2)$.)
This I do in the most naive way, sorting the data into increasing $\overline q^2$ and just 
taking successive differences. This rapidly becomes noisy at bigger $\overline q^2$,
so I arbitrarily kept the ten smallest values, $\overline q < 1.24$, for each $m_q$. I plot the result
in Fig.~\ref{fig:xtransl}. The collapse of the data to zero with the fermion mass is obvious.
Presumably, a true ``new physics'' prediction would multiply $S_{eff}$ by an overall scale representing
the coupling constant of the dynamics to the W or Z boson.

\begin{figure}
\begin{center}
\includegraphics[width=0.6\textwidth,clip]{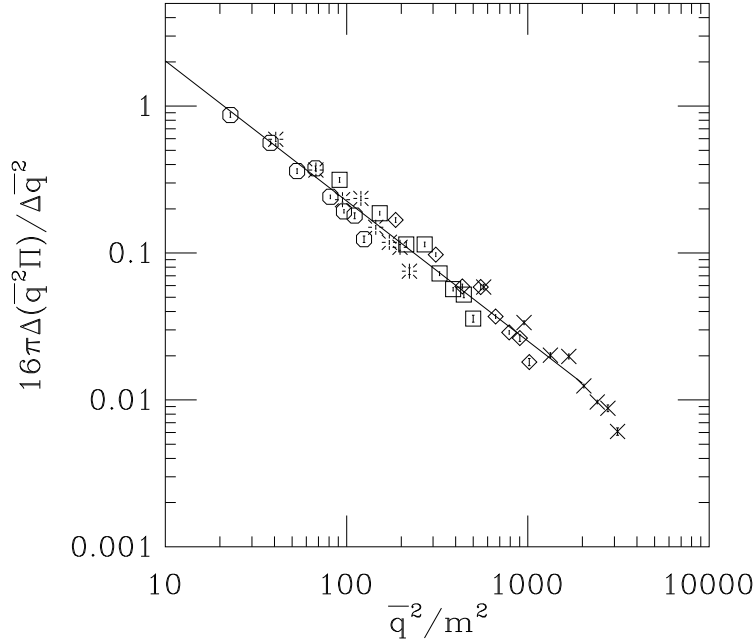}
\end{center}
\caption{Derivative of the sextet QCD correlator
$S_{eff}=16\pi\Delta(\overline q^2\Pi_T^{V-A})/\Delta \overline q^2$ plotted
as a function of the scaling combination $\overline q^2/m^2$.
Valence masses are (octagons) $am_q=0.10$, (bursts) $am_q=0.075$, (squares) $am_q=0.05$,
(diamonds) $am_q=0.035$. (crosses) $am_q=0.020$. The line is the straight power law
fit, Eq.~\protect{\ref{eq:pow}}.
\label{fig:xtranslres}}
\end{figure}

It is a worthwhile exercise to look for scaling behavior in the data. To do this, I replot the data
in Fig.~\ref{fig:xtranslres}. The data seems to fall on a scaling curve in terms of the
dimensionless ratio $\overline q^2/m^2$.  In free field theory, $S_{eff}=(48/\pi^2)(m^2/\overline q^2)$,
which seems to be a reasonable parameterization of the data. However, motivated
 by various didactic reviews \cite{didactic}   of the expected algebraic scaling
 behavior of conformal theories, I also
tried a little fit to
\bee
S_{eff} = a (\frac{m^2}{\overline q^2})^p.
\label{eq:pow}
\ee
The line is the result, $p=0.953(3)$. The uncorrelated chi-squared is 924 for 38 degrees of
freedom, but I should point out that the data is in fact strongly correlated.

Again, I remark that this is not the usual case studied in the technicolor literature, where the fermion mass is
often taken to be large compared to
$q$. Presumably to apply this result to some hidden sector beyond - Standard Model phenomenology,
one might take $q^2=m_Z^2$. (Or for another alternative, see Ref.~\cite{He:2001tp}.)

I believe that there are lessons to be drawn from this exercise.
First, because the system only has explicit chiral symmetry breaking at nonzero fermion mass,
$\Pi_T^{V-A}(q^2)$ does vanish at zero fermion mass. $\Pi_T^{V-A}(q^2)$ falls to zero for any value of the mass,
at any nonzero $q$.

Second, a simple parameterization of $\Pi_T^{V-A}(q^2)$ in terms of the (measured) properties of the
 lowest resonances fails, even though away from $m_q=0$ the theory is not conformal;
 it is a theory of resonances. Presumably, one must just include more of the
excited state spectrum in the sum. $S_{eff}$ shows power law scaling in terms of $m^2/\overline q^2$.

\section{Conclusions}
This paper was a qualitative investigation of a precision electroweak observable for
a lattice model of gauge fields and two flavors of symmetric representation fermions.
Before lattice simulations began, this model was one of many candidate theories for
walking technicolor  \cite{Sannino:2004qp,Hong:2004td,Dietrich:2006cm,Ryttov:2007sr}.
 At the bare parameters where I did my simulations, it does not
present the appearance of being a conventional walking technicolor model. However,
it does show conformal behavior ``for all practical purposes,'' that is, at the parameters
of the simulation the scale dependent gauge coupling runs very slowly \cite{DSS}.

I only asked simple questions, because the literature for these models is relatively sparse.
However, I discovered (as expected) that $\Pi_T^{V-A}(q^2)$ vanished for all $q$ as the quark
fermion mass vanished, and that it vanished at large $q$ for all values of the fermion mass.
The effective S-parameter is a scaling function of $m^2/q^2$.
I was surprised to discover that a phenomenologically popular parameterization of $\Pi_T^{V-A}(q^2)$ in terms of the
properties of the lowest resonances did not reproduce the data.

It might be, that all models are just different, and what I found for $\Pi_T^{V-A}(q^2)$
is not relevant to any other system. However, there are other would-be technicolor candidates
in the literature, which share some features with this model for regions of their bare parameter
spaces. $\Pi^T_{LR}$ in ``minimal walking technicolor'' (gauge group $SU(2)$ with a doublet of adjoint
representation fermions  \cite{Catterall:2007yx}) is an interesting target for a lattice study.

And of course, phenomenologists should expect more calculations of $\Pi_T^{V-A}(q^2)$,
done by  groups on models which are (hopefully) more attractive technicolor candidates -- perhaps
ones which are not so conformal as this one.

\begin{acknowledgments}
%%%%%%%%%%%%%%%%%%%%%%%%%%%%%%%%%%%%%%%%%%%%%%%%%%%%%%%%%%%%%%%%%%%%%%
This project was inspired by the lectures of R, Contino at the 2009 TASI summer
school \cite{Contino:2010rs}
I thank L.~Del Debbio for correspondence and 
B.~Svetitsky and Y.~Shamir for conversations.
I am grateful for the hospitality of the Niels Bohr International Academy during
the time I began this study.
This work was supported in part by the US Department of Energy.
%
%%%%%%%%%%%%%%%%%%%%%%%%%%%%%%%%%%%%%%%%%%%%%%%%%%%%%%%%%%%%%%%%%%%%
\end{acknowledgments}
%%%%%%%%%%%%%%%%%%%%%%%%%%%%%%%%%%%%%%%%%%%%%%%%%%%%%%%%%%%%%%%%%%%%%

\end{document}